\documentclass[12pt]{article}
\usepackage{axodraw,graphicx,amsbsy,latexsym,graphics}

\begin{document}

\newcommand{\sheptitle}
{SUSY, Inflation and the Origin of Matter in the
Universe\footnote{This is a review article to be published in J.Phys.G.
Based on a plenary talk given by S.F.K. at the IPPP BSM Workshop,
Durham, UK 6-11th May 2001.}}

\newcommand{\shepauthor}
{S. F. King and D. A. J. Rayner}

\newcommand{\shepaddress}
{Department of Physics and Astronomy,
University of Southampton, Southampton, SO17 1BJ, U.K.}

\newcommand{\shepabstract}
{We consider the Standard Models of particle physics and hot big bang 
cosmology, and review the theoretical and experimental motivations for 
extending these models to include supersymmetry and inflation.  An obvious
extension would be to unite these two models into a single 
all-encompassing theory.  We identify a list of theoretical challenges that 
such a theory must address, which we illustrate with a simple model - a 
variant of the next-to-minimal supersymmetric Standard Model - that addresses
these challenges.}

\begin{titlepage}
\begin{flushright}
SHEP 01-26
\end{flushright}
\vspace{0.5in}
\begin{center}
{\Large{\bf \sheptitle}}
\vspace{0.5in}
\bigskip \\ \shepauthor \\ \mbox{} \\ {\it \shepaddress} \\
\vspace{0.5in}
{\bf Abstract} \bigskip \end{center} \setcounter{page}{0}
\shepabstract
\end{titlepage}

\section{A Tale of Two Standard Models}  \label{sec:intro}

The Standard Model (SM) of particle physics provides a description of the 
fundamental particles and forces present in Nature.  It is expressed as a 
quantum field theory and combines quantum mechanics with special relativity
into a single consistent framework.  Local {\it gauge} symmetry is an 
essential ingredient, and the combined gauge group 
$SU(3)_{C} \otimes SU(2)_{L} \otimes U(1)_{Y}$ correctly describes 
electromagnetism and the strong and weak nuclear forces.  It can account
for the observed low-energy phenomena such as the infinite range of the 
electromagnetic force and radioactive decay of unstable nuclei in terms of
force-mediating quanta.  The model has been in place for over 30 years and has
been rigorously tested by experiments at high-energy particle accelerators.
There are many theoretical reasons to believe in a deeper theory - such as 
supersymmetry (SUSY) - but it is only in the last few years that new 
experiments have been able to probe physics 
beyond the Standard Model (BSM), e.g. massive neutrinos
and neutrino oscillations~\cite{neutrino}, $g_{\mu}-2$ measurements~\cite{g-2}
and even the recent Higgs candidate at LEP~\cite{lep}.   

There is a similar situation in cosmology where the hot big bang (HBB) 
Standard Model can account for the evolution of the early universe for cosmic
times $t \ge t_{P} \sim 10^{-44}s$ following the big bang, where quantum 
gravity effects are negligible.  The model
was developed after the two important discoveries of the cosmic microwave 
background (CMB) radiation and the Hubble expansion of the universe.
Among its many successes, the HBB paradigm can explain nucleosynthesis and 
reproduce the observed abundances of light elements; and predict a blackbody
CMB spectrum with the correct temperature of $T_{CMB} \sim 3 \, K$.  However
it has a number of long-standing problems that either require severely 
fine-tuned initial conditions, or a new theory - such as inflation - that 
provide observationally-consistent solutions to these problems in a natural 
way.  Recently satellite and balloon-based experiments have yielded evidence 
that verify the theoretical problems and provide experimental constraints for
any extended cosmological model - e.g. COBE density/temperature 
fluctuations~\cite{cobe}, and the BOOMERANG~\cite{boomerang}, 
MAXIMA~\cite{maxima} and DASI~\cite{dasi} observations of the angular 
power spectrum\footnote{These observations 
simultaneously provide information about the matter density and 
curvature of the universe today.}.

It would be desirable to combine the two Standard Models (and their
extensions) within a single all-encompassing theory - a supersymmetric 
inflationary model - that provides solutions for the
long-standing problems in each SM separately, and is also highly predictive
with fewer free parameters~\cite{king2}.  For example, such a theory may 
eventually unite string theory with cosmology~\cite{stringcosmo} since 
superstrings offer the best way of unifying all four fundamental forces in a 
mathematically consistent way.

The layout of the remainder of this review is as follows.  In section 
\ref{sec:whybsm} we discuss how supersymmetry and inflation solve the
theoretical and experimental problems of the Standard Models of particle 
physics and cosmology.  In section \ref{sec:ippp} we introduce the notion of 
an all-embracing theory that combines supersymmetry and inflation into a 
single unified framework.  Section \ref{sec:challenges} lists the challenges 
that a supersymmetric inflationary model must address, which  we illustrate 
with a well-studied example in section \ref{sec:phinmssm} - a variant of the 
next-to-minimal supersymmetric Standard Model (NMSSM).  Section 
\ref{sec:conc} concludes the review.

\section{Why go beyond the Standard Models?}  \label{sec:whybsm}

In this section we will show how low-energy SUSY and inflation tackle
the problems present in the Standard Models of particle physics and
cosmology.  There are many good references in the literature with further 
details~\cite{susy,cosmo}.
 
\subsection{Beyond the Particle Physics Standard Model - Supersymmetry}
 \label{sec:ppbsm}

The Standard Model of particle physics is a quantum field theory that unites
the two great successes of twentieth century physics - quantum mechanics and
special relativity.  It describes the fundamental forces and particles of 
Nature in terms of the local gauge group $SU(3)_{C} \otimes SU(2)_{L} \otimes 
U(1)_{Y}$, where force-carrying bosons mediate interactions between elementary
matter fermions and particles generate masses via their coupling to the Higgs 
boson.  This model has survived rigorous experimental tests at high-energy 
particle accelerators for over 30 years, but recently experiments have
begun to observe hints of new physics that cannot be explained by the Standard
Model\cite{neutrino,g-2,lep}.  These experimental anomalies support the many
theoretical reasons - such as the hierarchy and gauge coupling unification 
problem - which suggest that a new extended model is required.
\begin{figure}[h]
 \begin{center}
  \scalebox{0.8}{
    \begin{picture}(550,165)(0,30) 
        \DashLine(5,140)(55,140){6}
        \DashLine(115,140)(165,140){6}
        \ArrowArc(85,140)(30,0,180)
        \ArrowArc(85,140)(30,180,360)
        \Vertex(55,140){2}
        \Vertex(115,140){2}
        \Text(0,140)[r]{{\large $H_{u}$}}
        \Text(170,140)[l]{{\large $H_{u}$}}
        \Text(-20,165)[r]{{\Large (a)}}
        \Text(85,175)[b]{{\large $t_{L}$}}
        \Text(85,115)[b]{{\large $t_{R}$}}
        \Text(575,150)[r]{{\large $ \delta m_{H_{u}}^{2} \approx 
         \frac{y_{t}^{2}}{16 \pi^{2}} \left[ -2\Lambda_{UV}^{2}
          + 6 m_{t}^{2} \ln \left( \frac{\Lambda_{UV}}{m_{t}} \right) \right]
           + \ldots $}} 
        \Text(-20,70)[r]{{\Large (b)}}
        \DashLine(5,45)(55,45){6}
        \DashLine(115,45)(165,45){6}
        \ArrowArc(85,45)(30,0,180)
        \ArrowArc(85,45)(30,180,360)
        \Vertex(55,45){2}
        \Vertex(115,45){2}
        \Text(0,45)[r]{{\large $H_{u}$}}
        \Text(170,45)[l]{{\large $H_{u}$}}
        \Text(85,80)[b]{{\large $t_{L}$}}
        \Text(85,20)[b]{{\large $t_{R}$}}
        \Text(205,50)[]{{\LARGE $+$}}
        \DashLine(240,20)(310,20){6}
        \DashLine(310,20)(380,20){6}
        \DashCArc(310,50)(30,90,450){4}
        \Vertex(310,20){2}
        \Text(235,20)[r]{{\large $H_{u}$}}
        \Text(385,20)[l]{{\large $H_{u}$}}
        \Text(310,85)[b]{{\large $\tilde{t}_{L}$  ($\tilde{t}_{R}$)}}
        \Text(575,55)[r]{{\large $ \delta m_{H_{u}}^{2} \approx 
         - \frac{6 y_{t}^{2}}{16 \pi^{2}} m_{\tilde{t}}^{2} 
          \ln \left( \frac{\Lambda_{UV}}{m_{\tilde{t}}} \right) + \ldots$}} 
   \end{picture}
  }
 \end{center}
  \caption{{\small The dominant top (stop) 1-loop corrections to the Higgs
mass, where $y_{t}$ is the top(stop) Yukawa coupling. 
 In the absence of SUSY (a), only top loops contribute, and the 
radiative correction is found to be quadratically divergent in powers of the
ultraviolet momentum cutoff $\Lambda_{UV}$.  However, when stop loops are 
included (b), the quadratically divergent pieces cancel out to leave a softer
logarithmically divergent correction.  In the limit that SUSY is preserved 
($m_{\tilde{t}} = m_{t}$), there is an exact cancellation between the top
and stop 1-loop corrections to the Higgs mass.}}
    \label{fig:higgs}
\end{figure}

A leading candidate is supersymmetry - an underlying symmetry that
unites fermionic and bosonic degrees of freedom within the same 
{\it superfield} multiplets.  The minimal extension (MSSM) adds a 
fermion (boson) superpartner for each boson (fermion) particle in the Standard
Model\footnote{Notice that the (up-like) Higgs scalar obtains a spin-1/2 
Higgsino partner with identical gauge quantum numbers that leads to a gauge 
anomaly in the theory.  This requires that another (down-like) Higgs scalar 
and Higgsino must be 
added to cancel this anomaly.}.  SUSY combines internal and space-time 
Poincar\'{e} symmetries in a non-trivial way\footnote{The use of 
anti-commuting {\it Grassmann} variables evade the famous Coleman-Mandula 
No-Go theorem.}.  We know that a theory invariant with respect to these
symmetries can provide a realistic model of elementary particles and 
fundamental forces, so it is natural to want to unite internal and space-time
symmetries within {\it global} supersymmetry.  Notice that a {\it gauged} 
local
supersymmetry includes general coordinate transformations and necessarily 
incorporates a theory of gravity\footnote{Superstrings (so far) provide the 
only consistent quantum theory involving gravity, and supersymmetry is an 
essential ingredient.}.

SUSY solves the gauge hierarchy and naturalness problems by providing 
a symmetry that protects scalars (Higgs bosons) from acquiring masses of order
the underlying gravity (Planck) scale $M_{P}$
through radiative corrections.  Gauge fields are protected by an 
unbroken gauge invariance and fermions cannot acquire a large mass due to a
chiral symmetry.  As shown in figure \ref{fig:higgs},
SUSY stabilizes the puzzling ratio:
${\mbox m_{W}^{2}/M_{P}^{2} \simeq 10^{-34}}$ by contributing virtual 
sparticle loops for each particle loop that {\it soften} the quadratic 
divergence into a logarithmic divergence.  This avoids the unnecessary
fine-tuning problems, provided SUSY breaking (and consequently sparticle 
masses) appear around the TeV scale.  Supersymmetry also provides an
explanation for the mysterious Higgs mechanism.  Electroweak symmetry
breaking (EWSB) is triggered by radiative corrections to the Higgs
scalar masses, such that 1-loop corrections turn the up-like Higgs scalar
squared-mass negative at the origin.
\begin{figure}[h]
 \begin{center}
   \begin{picture}(300,120)(35,20)
     \hspace*{-2cm}
    \includegraphics[scale=0.45]{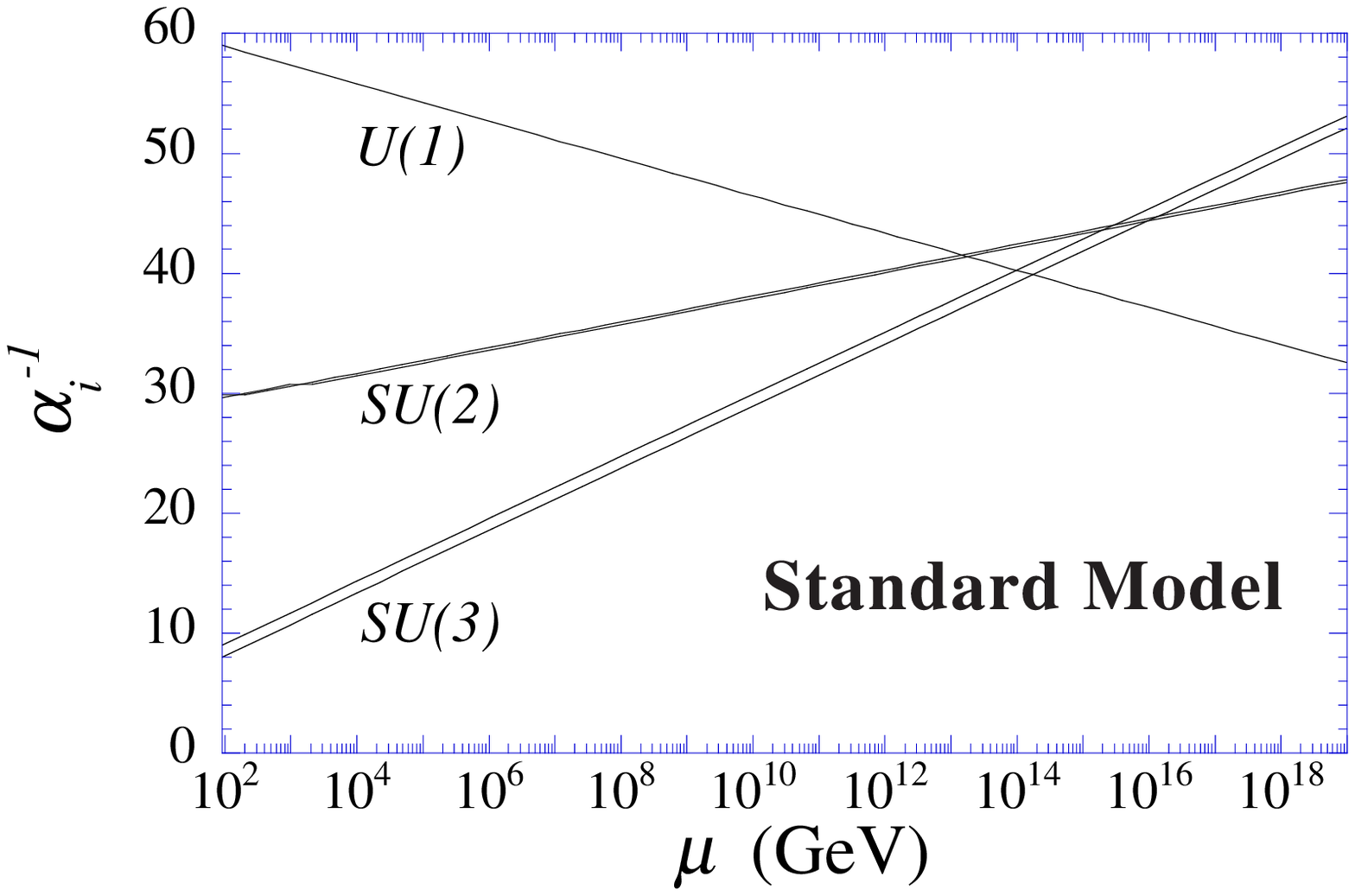}
     \hspace*{1cm}
    \includegraphics[scale=0.45]{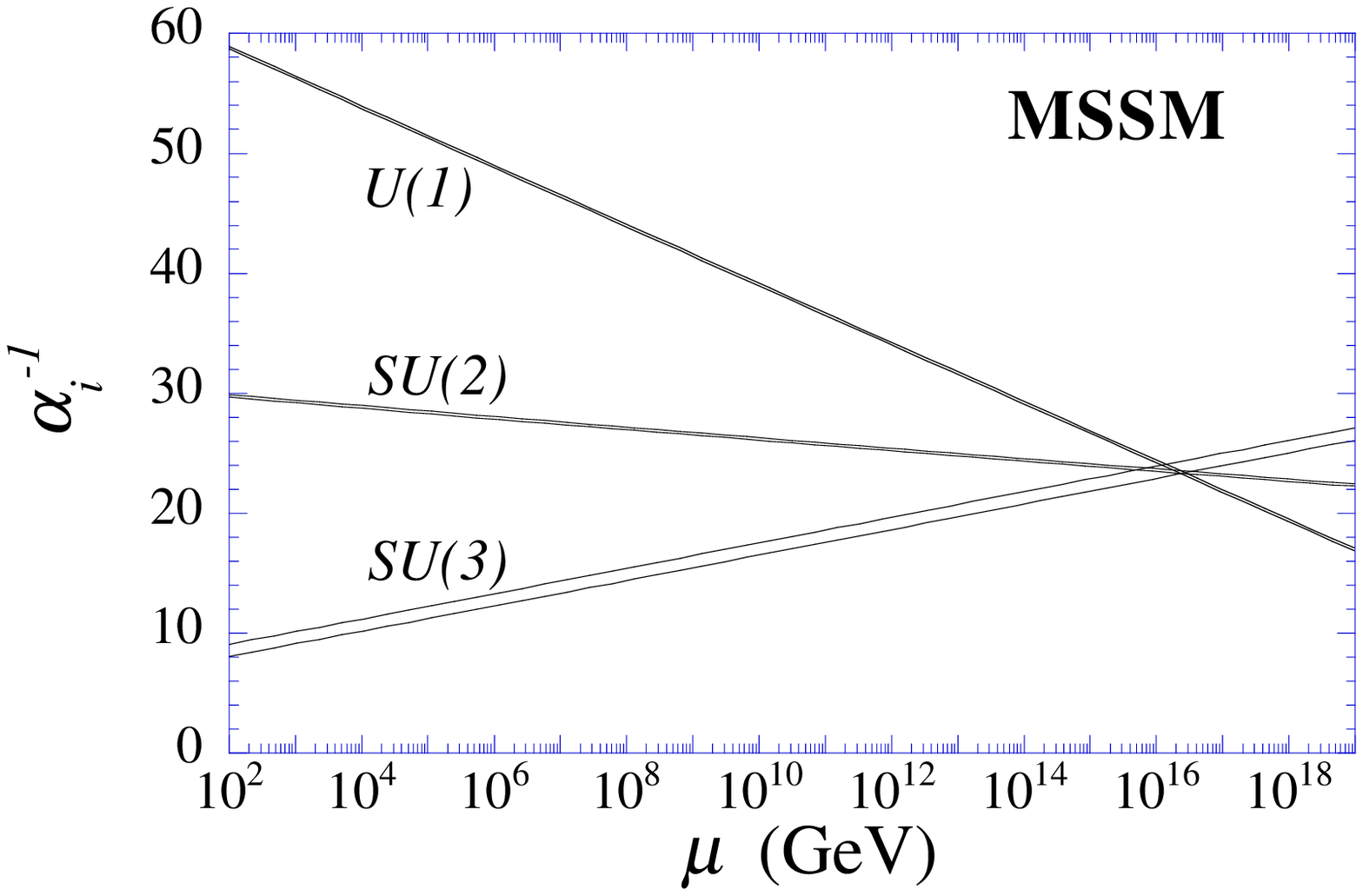}
   \end{picture}
 \end{center}
  \caption{{\small The renormalization group equation (RGE) running of gauge 
coupling constants $\alpha_{i}$ 
as a function of the renormalization scale $\mu$.  In the absence of SUSY, 
the three coupling constants do not meet at a single unified value.  However
in the MSSM, additional sparticle loops modify the evolution of the gauge 
couplings so that the coupling constants now meet at a unification scale of 
$M_{X} \simeq 10^{16}$ GeV.}}
    \label{fig:gcu}
\end{figure}

SUSY also modifies the gauge coupling renormalization group equation (RGE)
running by introducing higher-order loop corrections involving virtual SUSY 
partners as shown in figure \ref{fig:higgs}. 
The gauge coupling constants now meet at a scale 
${\mbox M_{X} \simeq 10^{16}}$GeV for $M_{SUSY} \sim 1$ TeV as shown in 
figure \ref{fig:gcu}.  The addition of supersymmetry to grand unified models
pushes the potentially dangerous proton decay rate {\it above} experimental 
lower bounds.  SUSY also offers a solution to the cold dark matter (CDM)
problem - the missing mass in the universe - in the form of the lightest
supersymmetric particle (LSP) that is very weakly-coupled and stable from 
decay due to a global R-parity conservation.  There are various models 
predicting the precise nature of the LSP, and neutralinos, gravitinos and 
axinos have all been considered\footnote{See L.Roszkowski's plenary talk at 
this meeting for a discussion.}.

However this is not to say that low-energy SUSY is complete.  There are still 
many unanswered questions, including the precise mechanism responsible for 
SUSY breaking\footnote{Many models have been proposed, but it will only be 
after we have observed supersymmetric sparticle spectra that we will be able
to identify the mechanism(s) responsible for SUSY breaking.} and the 
connection
of low-energy physics to the proposed underlying superstring theory.  However,
SUSY is an excellent candidate for TeV-scale physics and its predictions will
soon be tested at future accelerators.
  
\subsection{Beyond the Hot Big Bang Model - Inflation} 
 \label{sec:cosmobsm}

The HBB Standard Model of cosmology combines general relativity and classical
thermodynamics to describe the evolution of the universe for cosmic times 
$t \geq t_{P} \sim 10^{-44}$ s after the big bang, where quantum gravity 
effects are negligible.  The model can successfully reproduce the observed 
Hubble expansion of the universe; the existence of the cosmic microwave
background radiation with the correct temperature; and can also predict the 
relative abundances of light elements following nucleosynthesis.  In the 
simplest terms, the HBB model hypothesizes
that the universe exploded into existence (perhaps from quantum fluctuations)
as a microscopic ball with an unimaginably high temperature.  Following
unknown quantum effects, the universe contained a hot ``soup'' of massless 
particles (quarks, leptons, gauge bosons and Higgs fields) that rapidly 
cooled as it expanded in size.  As it cools, it undergoes a
series of phase transitions during which the four fundamental forces separate 
from a single unified interaction; massless quarks and leptons acquire masses 
as the Higgs mechanism breaks the electroweak symmetry; and quarks become 
bound together by the strong force to form hadrons.  Eventually the universe
cools down sufficiently that nucleosynthesis occurs, where protons and 
neutrons bind together as nuclei.  After the universe cools down further,
photons have insufficient energy to prevent electrons from binding to nuclei
to form neutral atoms.  The photons effectively decouple from matter and no
longer interact.  This is the epoch when the CMB radiation is formed, and 
begins to cool down to the temperature we observe today.
\begin{figure}[h!]
 \begin{center}
   \begin{picture}(300,170)(0,25)
    \includegraphics[scale=0.65]{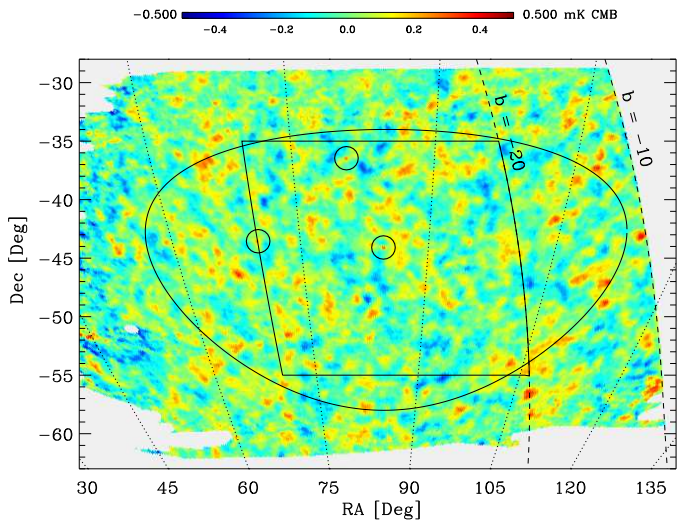}
  \end{picture}
 \end{center}
  \caption{{\small The sky map at 150 GHz, taken from BOOMERANG
~\cite{boomerang}, that shows the temperature anisotopies 
$\delta T/T \sim 10^{-5}$ in the cosmic microwave background radiation.  The 
location of three quasars are shown as circles.}}
    \label{fig:boomerang}
\end{figure}

Despite these theoretical successes, satellite and balloon-based experiments 
~\cite{boomerang}-\cite{dasi} have identified features that cannot be 
explained by the HBB model in a natural way without severe fine-tuning.  The 
BOOMERANG experiment~\cite{boomerang} observed a highly uniform CMB 
temperature in all directions in the sky as shown in figure 
\ref{fig:boomerang}.  This level of
uniformity requires that all of these regions were causally-connected when 
photons decoupled from matter, such that all regions equilibrated to a common
temperature.  Photons can only have travelled a finite distance, at the speed 
of light, since the CMB radiation was formed.  However this horizon is much
smaller than the size of the observable universe.  So, how did 
causally-unconnected regions of space achieve a uniform temperature to 1 part
in $10^{5}$?
A short period of exponential growth - or inflation - prior to the
power-law expansion of the HBB model, would solve this ``horizon problem''
since a small region of causally-connected (and thermalized) space could be
instantaneously stretched to a size greater than the observable universe.  
\begin{figure}[h]
 \begin{center}
   \begin{picture}(300,270)(0,25)
    \includegraphics[scale=0.55]{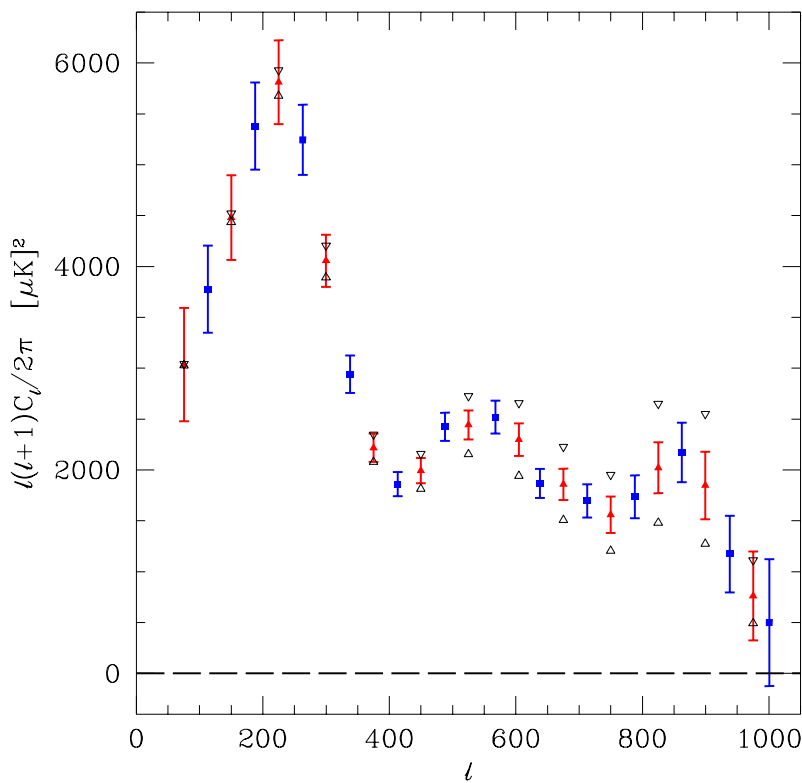}
  \end{picture}
 \end{center}
  \caption{{\small The angular power spectrum of the CMB, as measured at 150
GHz by BOOMERANG and taken from ref.~\cite{boomerang}.  The blue(square) and
red(triangle) points show the results of two independent analyses.  The basic
result is independent of binning.  The vertical error bars show the 
statistical $+$ sample variance errors on each point.  The location of the
first peak at $l \approx 200$ is consistent with a flat universe with total
density of unity $\Omega_{total} = 1 \pm 0.06$.  The presence of the smaller
amplitude acoustic peaks effectively rules out models involving topological
objects in the early universe such as textures and cosmic strings.  Notice 
that the details of the peaks are dependent on other cosmological quantities
such as the Hubble constant and baryon density.}}
    \label{fig:boomerangspectra}
\end{figure}

The universe is surprisingly uniform on cosmological scales, but we also know
that stars, planets and human-beings exist, which require density fluctuations
on smaller scales as shown in figure \ref{fig:boomerang} 
($\delta T/T = \delta \rho/\rho \sim 10^{-5}$).  How are such density 
fluctuations generated, while maintaining very large-scale uniformity? 
Inflation smoothes out any large-scale inhomogeneities in the initial 
conditions, but regenerates inhomogeneities by stretching quantum
fluctuations to an astronomical scale.  These fluctuations remain 
scale-invariant, and lead to the observed large-scale structure in the 
universe.

Figure \ref{fig:boomerangspectra} shows the angular power spectrum of the CMB
as measured at 150 GHz by BOOMERANG~\cite{boomerang}.  The location of the 
first peak at a multipole moment $l\approx 200$ corresponds to the angular
scale subtended by the Hubble radius at recombination, and is tied to the 
geometry of the universe\footnote{Photon paths diverge (converge) in a 
negatively (positively) curved universe which leads to a larger (smaller)
angular size compared to a flat universe with zero curvature. Negative 
(positive) curvature pushes the first peak to higher (lower) values of $l$.}. 
A flat universe has the first peak at $l \sim 200$, and the data provides the
best evidence that we live in a flat universe with a total density 
$\Omega_{total} \approx 1 \pm 0.06$.  Resolution of the second and third 
acoustic peaks in figure \ref{fig:boomerangspectra} provides very strong 
support for inflation, and effectively rules out models involving topological
objects such as textures and cosmic strings.  The data also supports a 
$\Lambda$CDM universe in which the energy density is dominated by dark energy
(possibly a cosmological constant $\Lambda$) and cold dark matter CDM.

\section{A Supersymmetric Inflationary Model}  \label{sec:ippp}

In this section we introduce the idea of an all-encompassing theory
that combines inflation with supersymmetry, and discuss the motivations for
such a model\footnote{For example, a supersymmetric model of inflation helps 
to keep the inflaton potential flat.}.  We also list
the issues that such a supersymmetric inflationary model must confront, and 
we give an explicit example that has already been well 
studied~\cite{phinmssm,mar}.  Recently, there has been a more detailed 
discussion of these issues in this supersymmetric inflationary model in 
ref.~\cite{king2}.

\subsection{Challenges for a Supersymmetric Inflationary model}  
 \label{sec:challenges}

Traditionally, the MSSM derives from a supersymmetric grand unified theory, 
with an increased unified gauge group such as $SU(5) , SU(5) \otimes U(1) , 
SO(10) , E_{6}$ or $E_{8}$.  This SUSY GUT in turn arises from an effective
supergravity model which is the low-energy realization of a superstring 
theory.  The phenomenologically desirable features in the low-energy theory 
should be derivable (in principle) from the underlying string theory.  
Unfortunately the lack of knowledge regarding the physical string vacuum state
and infinite class of allowed manifolds upon which the theory can be 
compactified, lead to a confusing ambiguity as to the precise details of the
superstring model.  However various ``bottom-up'' approaches to model-building
have identified ten important challenges that a supersymmetric inflationary 
model must be able to address.
\begin{enumerate}
 \item {\bf $\mu$-term} - the problematic ``Higgsino mass'' that mixes up and
down-like Higgsino fields in the superpotential.  Examples of possible 
solutions include the NMSSM where a gauge-singlet field is added to the MSSM
spectrum and generates a $\mu$-term after the singlet acquires a VEV. 
Alternatively the $\mu$-term may be forbidden in the superpotential by gauge 
invariance for models with larger gauge groups than the MSSM.  Instead it may
derive from the K\"{a}hler potential through the Giudice-Masiero 
mechanism~\cite{gm}.

 \item {\bf Strong CP problem} - the non-abelian gauge group $SU(3)_{C}$ 
describing the strong interaction allows a CP-violating lagrangian term, where
the amount of CP-violation is parametrized by an angle $\theta$.  However, 
experimental tests of the neutron electic-dipole moment show that strong
interactions preserve CP-symmetry to a very high accuracy, 
$| \theta | \le 10^{-12}$.  There is no explanation why $\theta$ is so small
without fine-tuning.  A popular solution imposes an approximate global, axial
$U(1)_{PQ}$ Peccei-Quinn symmetry~\cite{pq} that is broken at a very high 
energy scale
and allows the $\theta$ to be rotated away.  The breaking of $U(1)_{PQ}$
generates a pseudo-Goldstone boson (axion) that when combined with SUSY offers
a cold Dark Matter candidate (axino)~\cite{axino}.

 \item {\bf Right-handed neutrinos} - the fermions in the Standard Model are
divided into three families of quarks and leptons, where the left-handed 
fermions transform as doublets and the right-handed fields are singlets with
respect to $SU(2)_{L}$.  The absence of 
right-handed neutrinos in the Standard Model is inconsistent with the 
observation of (very small) neutrino masses~\cite{neutrino}.  If we add
right-handed neutrinos, we can form a gauge-invariant Yukawa term coupling 
neutrinos and a Higgs field together that will generate an electroweak-scale
Dirac mass after symmetry breaking.  The Standard Model gauge symmetry cannot
forbid the addition of 
a right-handed Majorana mass term at a high scale.  The see-saw mechanism can
now generate heavily suppressed neutrino masses consistent with experiment.
Note that grand unified models models based on the extended gauge groups of 
$SO(10)$ or $SU(4) \otimes SU(2)_{L} \otimes SU(2)_{R}$ naturally incorporate
right-handed neutrinos since quarks and leptons are unified within the same 
multiplets.

 \item {\bf SUSY breaking} - in the same way that the Higgs mechanism was the 
final piece of the Standard Model to be discovered, the precise mechanism 
responsible for SUSY breaking (and the splitting of SM particles and their 
superpartners) is one of the long-standing problems in supersymmetry.  A 
variety of viable mechanisms have been proposed - such as gravity, gauge, 
anomaly and gaugino mediation - that make predictions for the supersymmetric 
mass spectrum.  However we will only be able to identify the actual 
mechanism(s) responsible for breaking supersymmetry following the next 
generation of accelerators such as LHC and Tevatron Run II.

 \item {\bf Inflaton candidate} - inflation is driven by the vacuum energy of
a fundamental scalar field - the inflaton - that has so far eluded 
identification.  The NMSSM singlet, and even the two Higgs fields, have been
considered as candidates, but none of them provide a sufficiently flat
potential.  The conventional view is to invoke an additional scalar field (or
two such fields in the case of hybrid inflation\cite{hybridinf}) and assume 
that they arise from some deeper theory.

 \item {\bf Moduli problems} - big bang nucleosynthesis (BBN) places limits on
the time variation of the coupling constants in the SM.  In string theory, 
these couplings are related to the expectation values of moduli 
fields\footnote{Moduli and dilaton fields parametrize the geometry of the 
theory, especially ``flat directions'' in field space.} and can vary in time. 
Massive moduli fields are produced as non-thermal relics due to vacuum 
displacement.  They could make an embarrassingly large contribution to the 
critical density of the universe and must be removed~\cite{stringcosmo,dine}.
The moduli could decay to other particles, but this would 
destroy the successes of BBN.  Not only is moduli over-production a problem, 
but the dilaton must also be stabilized at a value that does not correspond 
to weak coupling in string theory\footnote{Dilaton stabilization has been 
recently discussed in the context of type I string theory~\cite{abel}.}.

 \item {\bf Gravitino problems} - in supergravity models constrained by Big 
Bang nucleosynthesis\footnote{SUSY is broken at an intermediate scale 
$\sim 10^{11} \, GeV$ in a hidden sector and communicated to the visible
sector via gravity mediation.}, the gravitino is predicted to have a mass
of $m_{3/2} \sim 10^{3} \, GeV$ which is comparable to the scale of SUSY
breaking in the visible sector.  The gravitino has very weak couplings 
(gravitational in origin) so that it decouples very early in the evolution of
the universe, leaving a large relic abundance after 
nucleosynthesis.  These slowly-decaying gravitinos ($\tau_{3/2} \sim 10^{3} s
> t_{BBN}$) produce a large number of high energy photons that can dilute 
baryons, and photodissociate nuclei to affect the agreement with the
observed abundances.  These relic gravitinos need to be removed 
somehow~\cite{mazumdar}.

 \item {\bf Baryogenesis\footnote{Notice that the next three challenges have 
no answer in the Standard Model, and strongly depend on the particular 
inflationary model.}} - the problem of the origin of baryons, specifically 
the source (and stabilization) of the observed baryon-antibaryon asymmetry. 
The inclusion of right-handed neutrinos can lead to baryogenesis via 
leptogenesis~\cite{lepto}.
Today baryons contribute $\Omega_{b} \sim 0.05$ to the energy density of the
universe, where recent observations suggest that a total energy density of 
unity is required.  This leaves the problem of the missing mass
in the universe that may occur in the form of cold Dark Matter (CDM) or as
``Dark Energy''.
 
 \item {\bf Cold Dark Matter candidates\footnote{This is the subject of 
L. Roszkowski's plenary talk at this meeting.}} - observations require the 
existence of so-far unobserved and very weakly-interacting particles - cold 
Dark Matter (CDM) - that contribute $\Omega_{CDM} \sim 0.3$ to the total 
energy density.  Supersymmetric extensions of the SM offer CDM candidate 
particles - gravitinos, neutalinos, axinos and neutrinos - that are stable and
therefore cannot decay into SM matter particles.

 \item {\bf Dark Energy problems} - there is still $\Omega_{\Lambda} \sim 
2/3$ of the total energy density in the universe that has not been identified.
The so-called ``Dark Energy'' density can either be time-independent 
(cosmological constant), or vary with time (quintessence).  However we need 
to understand why $\Omega_{\Lambda} \sim \Omega_{matter}$ now~\cite{arkani}. 
Recent work has considered how the dark energy density can be deduced from a
supersymmetric model of inflation using only the CMB temperature and Hubble
constant as input parameters~\cite{king2}. 

\end{enumerate}

\subsection{$\phi$NMSSM and Hybrid Inflation - an explicit example} 
 \label{sec:phinmssm}

We will now outline a model that one of us (S.F.K.) has worked on 
~\cite{phinmssm} - a variant of the 
NMSSM~\cite{nmssm} - that addresses the challenges set down in
section \ref{sec:challenges}.  There is a summary of the model in section 8.7 
of ref. ~\cite{riotto}.  We closely follow the recent analysis in 
ref.~\cite{king2}.

This variant of the NMSSM has the following superpotential terms involving the
standard Higgs doublets and two gauge singlet fields $\phi$ (inflaton) 
and $N$.
\begin{equation}
 W = \lambda N H_{u} H_{d} - k \phi N^{2}  \label{eq:super}
\end{equation}
Notice that the standard NMSSM is recovered if we replace the inflaton $\phi$ 
by N.  However this leads to the familiar domain wall problems arising from 
the discrete $Z_{3}$ symmetry.  In this new variant, the $Z_{3}$
becomes a global Peccei-Quinn $U(1)_{PQ}$ symmetry~\cite{pq} that is commonly 
invoked to solve the strong CP problem.  This symmetry is broken in the true 
vacuum by non-zero $\phi$ and N VEVS, where the axion is the pseudo-Goldstone
boson from the spontaneous symmetry breaking and constrains the size of the 
VEVS.  For the inflation model to work, axion physics require $\langle \phi
\rangle \sim \langle N \rangle \sim 10^{10}-10^{13}$ GeV.  

The $\mu$-term of the MSSM is identified as
\begin{equation}
 \mu \equiv \lambda \langle N \rangle \sim 10^{3} \, GeV
\end{equation}
which implies that $\lambda \sim 10^{-10}$ if $\langle N \rangle \sim 10^{13}$
GeV.  A model with such large VEVS gives an intermediate scale solution to the
$\mu$-problem, and will have collider signatures as discussed in 
ref.~\cite{spm}. The question remains why the coupling constants 
$\lambda , k$ appear to
be {\it unnaturally} small in comparison to the larger gauge-singlet VEVS at
$10^{13}$ GeV.  This problem has been addressed in ref.~\cite{kane} where 
such tiny couplings arise from non-renormalizable operators.

We can make the $\phi$-field real by a choice of the (approximately) massless 
axion field.  We will now regard $\phi$ and $N$ to be the real components of 
the complex singlets in what follows.
When we include soft SUSY breaking mass terms,
trilinear terms $A_{k}k \phi N^{2} + h.c.$ (for real $A_{k}$) and neglect the
$H_{u} H_{d}$ superpotential term, we have the following 
potential\footnote{Notice that since $\phi$ and $N$ are regarded as the real
components of the complex singlets, they must have the same overall factor of
$1/2$ in their mass terms.}:
\begin{eqnarray}
 V=V_{0} + k^{2} N^{4} + \frac{1}{2} m^{2}(\phi) N^{2} 
  + \frac{1}{2} m_{\phi}^{2} \phi^{2}
     \label{eq:potential} \\
 {\mathrm where} \hspace*{5mm} 
 m^{2}(\phi)= m_{N}^{2} + 4k^{2} \phi^{2} - 2k A_{k} \phi 
\end{eqnarray}
We can identify the various elements of the potential: $V_{0}$ arises
from some other sector of the theory, SUGRA for example, and dominates the
potential; the soft SUSY breaking parameters $A_{k}$ and $m_{N}$ are
generated through some
gravity-mediated mechanism with a generic value of ${\mathcal O}(TeV)$; and
$m_{\phi}$ comes from no-scale SUGRA, and
vanishes at the Planck scale\footnote{It is generated through radiative
corrections such that $m_{\phi}^{2} \sim -k^{2} A_{k}^{2} \sim 
-(100eV)^{2}$.}.
  
Note that the $N$-field is destabilized if $\phi$ lies between the values:
\begin{equation}
 \phi_{c}^{\pm} = \frac{A_{k}}{4k} \left( 1 \pm \sqrt{ 1 - 
  \frac{4 m_{N}^{2}}{A_{k}^{2}}} \right)   \label{eq:phicrit}
\end{equation}
where we are assuming that $4 m_{N}^{2} < A_{k}^{2}$ in the following 
analysis.
\begin{figure}[h]
 \begin{center}
   \begin{picture}(300,200)(0,20)
    \includegraphics[scale=0.5]{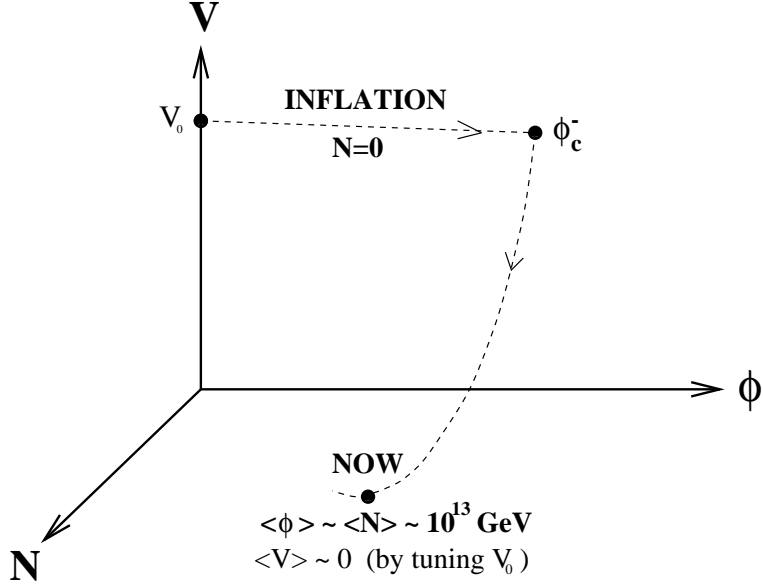}
  \end{picture}
 \end{center}
  \caption{{\small During (inverted hybrid) inflation the singlet $N$ is 
trapped at the origin and the 
inflaton rolls towards a critical value at $\phi_{c}^{-}$, whereupon
the potential acquires an instability and rolls down towards the true global
minimum (NOW) where the singlets have VEVS $\sim 10^{13} GeV$ and the
cosmological constant vanishes in agreement with observation.  Notice that 
$\phi$ and $N$ are the real components of the complex singlets.}}
    \label{fig:potential}
\end{figure}

In order to discuss inflation as illustrated by figure \ref{fig:potential},
we need to specify the sign of the inflaton mass
squared $m_{\phi}^{2}$.  If $m_{\phi}^{2} >0$ (hybrid inflation) then, for
$\phi > \phi_{c}^{+}$, $N$ will be driven to a local minimum 
(false vacuum) with $N=0$.  $\phi$ will roll towards the origin and 
$m_{\phi}^{2}$ will change 
signs and become negative for $\phi \approx \phi_{c}^{+}$.  Following this
sign change, the potential develops an instability in the $N=0$ direction 
and both singlets roll down towards the global minimum (true vacuum) at:
\begin{eqnarray}
 \langle \phi \rangle &=& \frac{A_{k}}{4k} \\
 \langle N \rangle &=& \frac{A_{k}}{2\sqrt{2} k} 
  \sqrt{ 1- \frac{4 m_{N}^{2}}{A_{k}^{2}}} 
 = \sqrt{2} \left| \phi_{c}^{\pm} - \langle \phi \rangle \right|
\end{eqnarray}
that signals the end of inflation.

However if $m_{\phi}^{2} < 0$ (inverted hybrid inflation), we suppose that
during inflation $\phi < \phi_{c}^{-}$, and the inflaton rolls away from the
origin, eventually reaching $\phi_{c}^{-}$ and ending inflation at the same
global mimimum as before.  Notice that the global (true vacuum) VEV 
$\langle \phi \rangle$ lies between $\phi_{c}^{-}$ and $\phi_{c}^{+}$, so
either hybrid or inverted hybrid inflation is possible\footnote{However the
radiative corrections to the inflaton mass actually give inverted hybrid 
inflation as shown in figure \ref{fig:potential}.} depending on the sign of 
the inflaton mass squared $m_{\phi}^{2}$.

We will also ignore the tiny effect of $m_{\phi}$ when we calculated the
true vacuum VEVS to obtain the following order of magnitude results:
\begin{eqnarray}
  A_{k} \sim k \phi_{c}^{\pm} \sim k \langle N \rangle 
   \sim k \langle \phi \rangle \sim 1 TeV
     \label{eq:ak}
\end{eqnarray}
For VEVS at the axion scale $\sim 10^{13} GeV$, we require that $k \sim
{\mathcal O}(10^{-10})$, and $\lambda$ must also take a similarly small value
since the combination $\lambda \langle N \rangle$ provide the 
$\mu$-parameter.
Notice that the SUGRA-derived potential contribution $V_{0}$ exactly cancels
with the other terms (by tuning) to provide agreement with the observed small
cosmological constant.  Thus we assume:
\begin{eqnarray}
 V(0) = - V(\langle \phi \rangle , \langle N \rangle )
  = k^{2} \langle N \rangle^{4} 
   = 4k^{2} (\phi_{c}^{\pm} - \langle \phi \rangle )^{4}
\end{eqnarray}
We may set $N=0$ during inflation, so the potential of 
eq.(\ref{eq:potential}) simplifies to:
\begin{eqnarray}
 V = V(0) + \frac{1}{2} m_{\phi}^{2} \phi^{2}
   \label{eq:potential2}
\end{eqnarray}

During inflation, the inflaton field $\phi$
is supposed to be on a region of the potential which satisfies the following
{\it flatness conditions}\footnote{See ref.~\cite{liddlelyth} for further 
details.}:
\begin{eqnarray} 
 \epsilon &\equiv& \frac{1}{2} \tilde{M}_{P}^{2} 
  \left( \frac{V'}{V} \right)^{2} \ll 1  \label{eq:slowepsilon} \\
 | \eta | &\equiv& \left| \frac{ \tilde{M}_{P}^{2} V''}{V} \right| \ll 1
  \label{eq:sloweta}
\end{eqnarray} 
where $V' (V'')$ are the first (second) derivatives of the potential, and
$\tilde{M}^{2}_{P} = M^{2}_{P}/8 \pi$ is the reduced Planck mass.
From eqs.(\ref{eq:slowepsilon},\ref{eq:sloweta}), the slow roll conditions 
are given by:
\begin{eqnarray}
 \epsilon_{{\mathcal N}} &=& \frac{M_{P}^{2} m_{\phi}^{4} 
  \phi_{{\mathcal N}}^{2}}{16 \pi V(0)^{2}} \ll 1 \\
 \left| \eta_{{\mathcal N}} \right| &=& \frac{M_{P}^{2} |m_{\phi}^{2}|}{8
  \pi V(0)} \ll 1
    \label{eq:slowroll}
\end{eqnarray}
where $\epsilon_{{\mathcal N}} , \eta_{{\mathcal N}}$ and 
$\phi_{{\mathcal N}}$ are evaluated around ${\mathcal N}=60$ 
e-folds before the end of inflation\footnote{The scale factors $a(t)$ before 
and after inflation are related by $a(t_{after})/a(t_{end}) 
= e^{{\mathcal N}}$, where ${\mathcal N}$ is called the number of 
``e-folds''.} and $V(0)$ is the dominant term in 
eq.(\ref{eq:potential2}) during inflation, $\phi_{{\mathcal N}} 
 = \phi_{c}^{\pm} e^{\eta {\mathcal N}} \approx \phi_{c}^{\pm}$.
The height of the potential during inflation is approximately constant and
given by:
\begin{equation}
 V_{0}^{1/4} \sim k^{1/2} \langle N \rangle \sim 10^{8} GeV
   \label{eq:height}
\end{equation}

We need to check that we can reproduce the correct level of density 
perturbation - responsible for the large scale structure in the universe -
according to the COBE anisotropy measurements, where the spectrum of 
perturbations is given by\cite{liddle}:
\begin{eqnarray}
 \delta_{H}^{2} = \frac{32 V(0)}{75 M_{P}^{4} \epsilon_{{\mathcal N}}}
   \label{eq:cobedh}
\end{eqnarray}
with the COBE value, $\delta_{H} = 1.95 \times 10^{-5}$ ~\cite{bunn}.
Writing $\phi^{\pm}_{c} \sim \phi_{c}$ and combining eqs.(\ref{eq:ak}, 
\ref{eq:slowepsilon}, \ref{eq:height}, \ref{eq:cobedh}), we obtain the order
of magnitude constraint:
\begin{eqnarray}
 | k m_{\phi} | \simeq 8 \left( \frac{8\pi}{75} \right)^{1/4} 
  \delta_{H}^{-1/2} \frac{ (k \phi_{c})^{5/2}}{M_{P}^{3/2}} 
 \simeq 10^{-18} \, GeV \, \left( \frac{k \phi_{c} }{1 \, TeV} \right)^{5/2}
\end{eqnarray} 
which is adequate to broadly satisfy the slow-roll conditions of 
eqs.(\ref{eq:slowepsilon},\ref{eq:sloweta})
\begin{eqnarray}
 | \eta_{{\mathcal N}} | &\simeq& \frac{ M_{P}^{2}}{8 \pi}
  \frac{ | k m_{\phi} |^{2} }{ \left( \sqrt{2} k \phi_{c} \right)^{4} }
   \sim 10^{-12} , 
     \label{eq:sloweta2} \\
 \epsilon_{{\mathcal N}} &\sim& \frac{ M_{P}^{2}}{16 \pi}
  \frac{ | k m_{\phi} |^{4} }{ \left( \sqrt{2} k \phi_{c} \right)^{8} } 
   \phi_{{\mathcal N}}^{2} \sim 4 \pi 
    \frac{ \phi_{{\mathcal N}}^{2}}{M_{P}^{2}} \eta_{{\mathcal N}}^{2}
      \label{eq:slowepsilon2}
\end{eqnarray}
This (inverted) hybrid model predicts a very flat spectrum of density 
perturbations, with no appreciable deviation in the spectral index,
$n=1+2\eta - 6\epsilon$ from unity which is consistent with observations and
predictions from an $n=1$ scale-invariant Harrison-Zel'dovich spectrum.
Notice that the COBE results require the product $|k m_{\phi}|$ to be 
extremely small, which implies that the inflaton mass is in the electronvolt 
range ($m_{\phi} \sim eV$) when we take $k \sim 10^{-10}$ which is motivated 
by axion physics as discussed earlier.

Inflation ends with the singlets $\phi,N$ oscillating about their global 
minimum.  Although the final reheating temperature is estimated to be of order
1 GeV~\cite{phinmssm}, during the reheating process the effective temperature
of the universe (as determined by the radiation density) can be viewed as 
rapidly rising to $V_{0}^{1/4} =k^{1/2} \langle N \rangle \sim 10^{8} GeV$
then slowly falling to the final reheat temperature during the reheating 
process~\cite{mar}.  This reheating gives entropy to the universe.  
Non-perturbative effects can produce particles with masses up to the potential
height, i.e. $m \leq V_{0}^{1/4} \sim 10^{8} GeV$ (preheating)
~\cite{preheating}.  We can check that problematic axions and gravitinos are 
not over produced~\cite{mazumdar,sanderson}.  
The superpotential is modified since Higgses and right-handed sneutrinos 
$\tilde{\nu}_{R}$ are copiously produced during this preheating phase via the
couplings $\lambda$ and $k$ to the oscillating inflaton fields.
\begin{equation}
 W = \lambda N H_{u} H_{d} - k \phi N^{2}
  + Y_{\nu} L \cdot H_{u} \nu_{R} + M \nu_{R} \nu_{R}
\end{equation}
These additional superpotential terms solve the problem of non-zero neutrino 
masses.  The right-handed neutrinos are SM gauge-singlets and so a heavy 
Majorana mass term can be added at a high energy scale.  Neutrino Dirac 
masses are then generated by electroweak symmetry breaking from the Yukawa
 coupling terms in the superpotential. The see-saw mechanism generates two
mass eigenvalues - one is very large (above the reach of current detection) - 
and the other is very light and therefore consistent with the recent
experimental constraints.

Now consider the origin of matter in the universe\footnote{We will only give
a summary of the results since a detailed discussion is given in 
ref.~\cite{mar}.}. Baryons originate from sleptogenesis~\cite{lepto} via the 
out-of-equilibrium decay of right-handed sneutrinos ($\tilde{\nu}_{R}$) and 
Higgses that violate lepton number (and hence $B-L$) asymmetry before 
subsequently converting into baryon number asymmetry through sphaleron 
interactions.  From the perspective of inflation, the conventional 
leptogenesis picture will change if the reheat temperature is below
the mass of the lightest right-handed neutrino. 
Notice that, unlike the usual hot big bang scenario, the 
out-of-equilibrium condition is automatically satisfied during reheating, and 
the production mechanism of right-handed neutrinos is totally different and 
due to direct or indirect couplings to the inflaton field.

In the standard hot big bang scenario, the baryon asymmetry is given by:
\begin{equation}
 Y_{b} \sim \frac{d \epsilon}{g^{\ast}}
\end{equation}
where $\epsilon$ is the lepton number asymmetry produced in the decay of the 
lightest right-handed neutrino
\begin{eqnarray}
 \epsilon = \frac{ \Gamma \left( \tilde{\nu}_{R} \rightarrow \tilde{l} + H_{u}
  \right) 
 - \Gamma \left( \tilde{\nu}_{R} \rightarrow  \overline{\tilde{l}} + 
  \overline{H}_{u} \right) }{ 
 \Gamma \left( \tilde{\nu}_{R} \rightarrow \tilde{l} + H_{u} \right) 
 + \Gamma \left( \tilde{\nu}_{R} \rightarrow  \overline{\tilde{l}} + 
  \overline{H}_{u} \right) }  
    \label{eq:epsilon}
\end{eqnarray}
$g^{\ast}$ counts the effective number of degrees of freedom\footnote{For
the SM, $g^{\ast}=106.75$, and for the MSSM $g^{\ast}=228.75$.}, and $d$ is 
the dilution factor.  However for the non-standard leptogenesis picture 
outlined above, the baryon asymmetry is given by:
\begin{equation}
 Y_{b} \sim \frac{ \gamma \epsilon (c V_{0})^{1/4} }{M_{1}}
\end{equation}
where $c$ is the fraction of the total vacuum energy density converted into 
right handed neutrinos (mass $M_{1}$) due to preheating, and $\gamma$ accounts
for dilution due to entropy production during reheating.

There are two primary contributions to the cold Dark Matter candidate 
particles, either supersymmetric partners or the Peccei-Quinn axion.
\begin{itemize}
 \item Neutralino~\cite{neutralino} / singlino~\cite{spm}
  / inflatino~\cite{inflatino} / axino~\cite{axino} -
the Higgs bosons can decay into radiation and a neutralino $H_{u},H_{d}
\rightarrow \gamma + \tilde{\chi}^{0}$, where the neutralino can subsequently
decay into an inflatino $\tilde{\phi}$, singlino $\tilde{N}$ or axino
$\tilde{a} \sim \alpha \tilde{\phi} + \beta \tilde{N}$, provided that they are
lighter than the neutralino $\tilde{\chi}^{0}$.
 \item Axions - relativistic axions are produced during preheating and so they
are red-shifted away; non-relativistic axions are generated at the QCD scale
by the misalignment mechanism, and make CDM candidates.
\end{itemize}
\begin{figure}[h]
 \begin{center}
   \begin{picture}(300,240)(0,20)
    \includegraphics[scale=0.6]{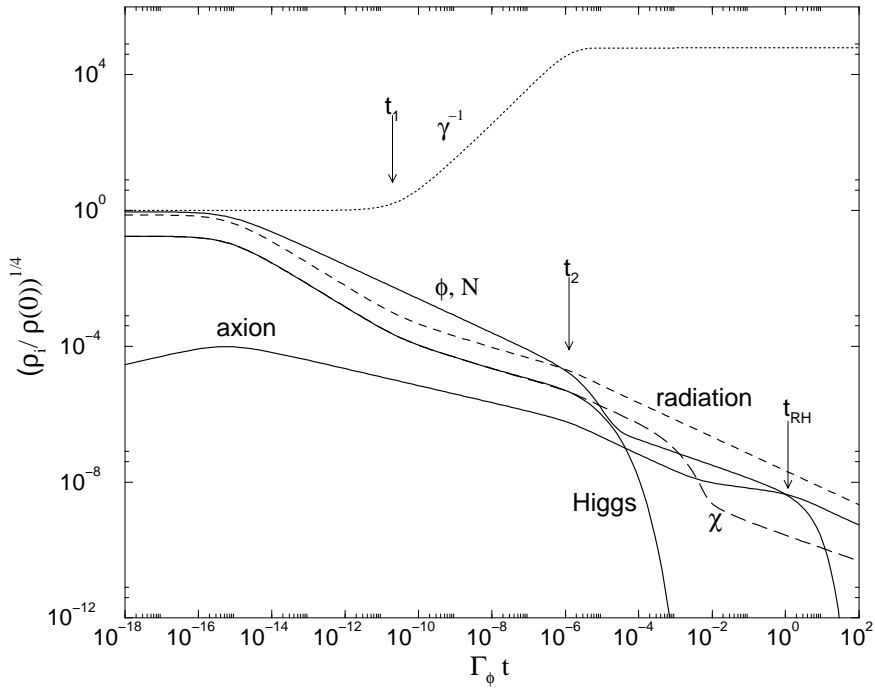}
   \end{picture}
 \end{center}
  \caption{{\small Evolution of the energy densities of the singlets 
($\phi,N$), axions, Higgses, radiation (dashed line) and neutralinos 
(long-dashed line).  The full analysis and figure are taken from 
ref.~\cite{mar}.}}
    \label{fig:rho}
\end{figure}
The time evolution of the various particle densities can be determined 
by solving a series of coupled Boltzmann equations~\cite{mar} for a 
particular choice of parameters as shown in figure \ref{fig:rho}.  In 
principle, we can calculate the densities of neutralinos, radiation, 
relativistic axions and baryons at reheating time $t_{RH}$ (defined as the 
time at which the oscillating singlet energy density rapidly decayed to 
zero~\cite{mar}) which represents the start of the hot big bang.

The important point to emphasize for a given model is that, at this time 
$t_{RH}$, the Boltzmann equations allow us to calculate the energy densities
of matter and radiation (at $t_{RH}$).  As pointed out recently~\cite{king2},
this allows us to calculate the dark energy density (in principle) without 
having a specific dark energy model in mind, but only inputting the
CMB temperature $T_{CMB}$ and Hubble constant.

Finally we will mention a few important points about SUGRA, where the 
K\"{a}hler potential can be split into separate K\"{a}hler potential and 
superpotential terms that are functions of the dilaton ($S$), an overall 
moduli ($T$), inflaton ($\phi$) and gauge singlet ($N$) fields.  These 
functions include non-perturbative terms to stabilize the dilaton and moduli 
potentials.
\begin{eqnarray}
 G &=& K + \ln |W|^{2} \\
 K &=& -3 \ln (\rho) + \frac{ \beta_{np}}{\rho^{3}} - \ln ( S + S^{\ast} )
  + \hat{K}_{np}(S) \\
 W &=& -k \phi N^{2} + \Lambda^{3} e^{-S/b_{0}} + \ldots
  \label{eq:kahler}
\end{eqnarray}
where $\rho = (T + T^{\ast}) - \phi^{\ast} \phi - N^{\ast} N$, and we assume 
an overall modulus $T$.  Notice that $\beta_{np}/\rho^{3}$, $\hat{K}_{np}(S)$ 
and $\Lambda^{3} e^{-S/b_{0}}$ arise through non-perturbative mechanisms.

Notice that eq.(\ref{eq:kahler}) has a no-scale structure with 
$m_{\phi}=m_{N}=0$ at tree-level.  As mentioned earlier, the cosmological 
constant can be tuned to zero by an appropriate choice of $V_{0}=|F_{S}|^{2}
+ |F_{\rho}|^{2} \sim (10^{8} \, GeV)^{4}$.

During inflation, the dilaton $S$
and ``moduli'' $\rho$ are stabilized at their respective minima since as the
inflaton $\phi$ rolls (and $N=0$), the overall modulus field $T$ adjusts to 
keep the combination $\rho = (T + T^{\ast}) - \phi^{\ast} \phi$ fixed.
After inflation, $S$ and $\rho$ only shift by $\sim 10^{-10}$, and so there is
no moduli problem.  However, it is important to clarify the connection with
string theory, e.g. stabilization of the dilaton potential in type I string
models~\cite{abel}.

\section{Final Thoughts}  \label{sec:conc}

We will soon see considerable progress in cosmology and supersymmetric
particle physics due to the observations of the Map and Planck explorer 
satellites, and the Tevatron and LHC accelerators.  These experiments will
accurately measure the fundamental parameters such as the
abundances $\Omega_{b},\Omega_{CDM}$ and $\Omega_{\Lambda}$ present in the 
universe and the supersymmetric mixing angles and soft parameters.  
However {\it ab initio} predictions of these parameters are difficult to 
obtain, but within the framework of an all-embracing supersymmetric 
inflationary model there will be fewer variables (and more predictivity) 
since the same parameters control both inflation {\it and} collider 
physics~\cite{king2}.

The NMSSM variant discussed in section \ref{sec:phinmssm} is an example of 
such a model which addresses the ten theoretical challenges outlined in 
section \ref{sec:challenges}.  However we admit that it is a long way from
being able to make accurate predictions for $\Omega_{b},\Omega_{CDM}$ and 
$\Omega_{\Lambda}$.  In particular there also needs to be an explanation why
$\rho_{\Lambda}^{1/4} \sim M_{W}^{2}/M_{P}$.  However, this model is a step
towards more realistic supersymmetric inflationary models that may also
incorporate superstring theory.

\newpage
\begin{center}
{\bf \large Acknowledgements}
\end{center}
S.K. and D.R. would like to thank PPARC for a Senior Fellowship
and a Studentship.  We would both like to thank the IPPP for their 
hospitality and support.  We would also like to thank all of the
organisers at the IPPP for arranging such a stimulating BSM meeting.
We would like to thank James Hockings for reading the manuscript.


\end{document}